\documentclass[sigconf]{acmart}

\usepackage{booktabs} 
\usepackage{float}
\usepackage{pstricks,pst-node,pst-tree}
\usepackage{graphicx}
\usepackage{caption}
\usepackage{subcaption}
\usepackage{fancybox}
\usepackage{tikz}
\usepackage{tikz-qtree}
\graphicspath{ {./images/} }

\setcopyright{acmcopyright}

\setcopyright{acmlicensed}
\acmConference[DAPA '19]{DAPA 2019 WSDM Workshop on Deep matching in Practical Applications}{February 15th, 2019}{Melbourne, Australia} \acmYear{2019} \copyrightyear{2019} \acmDOI{} \acmISBN{} \acmPrice{}

\begin{document}
\title{Approximating Categorical Similarity in Sponsored Search Relevance}

\author{Hiba Ahsan}
\affiliation{
  \institution{Microsoft India Development Center}
}
\email{hiahsan@microsoft.com}

\author{Rahul Agrawal}
\affiliation{
	\institution{Microsoft India Development Center}
}
\email{rahulagr@microsoft.com}


\begin{abstract}
 Sponsored Search is a major source of revenue for web search engines. Since sponsored search follows a pay-per-click model, showing relevant ads for receiving clicks is crucial. Matching categories of a query and its ad candidates have been explored in modeling relevance of query-ad pairs. The approach involves matching cached categories of queries seen in the past to categories of candidate ads. Since queries have a heavy tail distribution, the approach has limited coverage. In this work, we propose approximating categorical similarity of a query-ad pairs using neural networks, particularly CLSM \cite{CLSM}. Embedding of a query (or document) is generated using its tri-letter representation which allows coverage of tail queries. Offline experiments of incorporating this feature as opposed to using the categories directly show a 5.23\% improvement in AUC ROC. A/B testing results show an improvement of 8.2\% in relevance.
\end{abstract}

%
%
%

\keywords{Sponsored Search, Categorization, Neural Networks}

\maketitle

\section{Introduction}
\label{intro}
Sponsored search is a major source of revenue for web search engines. It displays ads as results to queries. Advertisers bid on keywords on which they want their ads to trigger and pay when their ads receive clicks. The search engine aims to generate revenue by driving clicks on relevant ads.
Advertisers can specify how they want queries to be matched to their bidded keyword. There are three types of matches: Exact, Phrase and Broad. An Exact match matches the query exactly as it is to the bidded keyword. A Phrase match requires the bidded keyword to be a phrase within the query. A Broad match has no lexical based rule and only requires the ad to be relevant to the query.
Once candidate ads for a query have been generated through various techniques \cite{ContextEmbed, FastText}, the relevance of the candidates are determined and irrelevant ads are filtered. This is important since irrelevant ads cause dissatisfaction among users, advertisers and affect the engine's goal of generating revenue.

Commercial web search engines model relevance using text-based features, both lexical \cite{RoleRev} and semantic \cite{NGram}. Click-based features are avoided since an ad receiving a click does not ensure its relevance to the query \cite{RoleRev}. Previous works have explored using categorization of queries and ads in modeling relevance \cite{AdvIntent}.

Categorization of a query (ad) helps capture its intent. The query (ad) is classified into one or more predefined categories such as \textit{Restaurants} or \textit{Tourism}. Computing the similarity of a query-ad pair in terms of their categories helps pick pairs with similar intent and eliminate arbitrary matches.
To be able to compute the categorical similarity of a query with its retrieved ads at runtime, categories of all ads available in the ad corpus are computed offline and indexed. The same is done for the most frequent queries in the search logs. For a query-ad pair at runtime, the categories of the query and ad are looked up in the index. But this approach has poor coverage. Queries have a heavy tail distribution \cite{Tail}. Using cached categories limits us to only the head and torso queries which make up a very small portion of the query distribution. 
In this paper, we propose approximating the categorical similarity of a query-ad pair to address the above issue. Instead of explicitly extracting categories of the query and the ad and then computing the similarity in terms of the categories, we approximate the similarity using a deep neural network model (CLSM).  Since the model is lightweight, the computation can be done at runtime. The model is robust to out-of-vocabulary queries and hence, covers the tail queries as well.

\section{Related Work}
\subsection{Semantic representation learning}
Semantic representation learning map words to a low dimensional vector space where representations of semantically similar words lay close to each other. Work in this area include LSA \cite{LSA}, pLSI \cite{pLSI} and LDA \cite{LDA}. Neural language models learn embeddings using neural networks \cite{CBOW, SNGS}. Neural semantic embeddings have been applied in the sponsored search domain for retrieval \cite{SemMatch, WebClick} as well as relevance \cite{NGram}.
 
\subsection{Relevance in Sponsored Search}
While the problem of optimizing the click through rate of query-ad pairs is a well-studied problem, improving relevance in sponsored search is relatively less investigated. \cite{ImproveRel} propose using translation model scores trained on click logs and text overlap features in modeling relevance. \cite{RoleRev} argue that a high click through rate (CTR) does not necessarily assure high relevance. They propose a relevance model trained on text-based features including token overlap, Jaccard distance between query and various ad components and LSI based query-ad similarity. \cite{AdvIntent} leverage the search engine to understand the advertiser intent by querying the keyword and then using derived features in their relevance model. \cite{NGram} propose improving relevance by using a query n-gram embedding as a semantic feature in the query-ad relevance model.

\subsection{Categorization in Search}
Exploiting categories for retrieving better matches in web and sponsored search is common. Query expansion techniques have used query categorization to improve search results and ad matching \cite{Class}. \cite{Boulder} develop an ontology to group offline purchase intents for recommending offline retail locations. \cite{Bounce} use the categories of the ad keyword, creative and landing page to predict the bounce rate of an ad. \cite{AdvIntent} compute the semantic similarity of a query-ad pair in terms of the categories of their respective organic search results.

\section{Matching categories of query and ad}
Previous methods that have computed the categorical similarity of a query and an ad have either directly used their top k categories as features [5] or have used simple overlap techniques [6]. As discussed in section \ref{intro}, these methods require caching of categories and are limited in their inability to categorize tail queries at runtime. To overcome these limitations, we model the categorical similarity using CLSM \cite{CLSM}. We choose CLSM because:
\begin{enumerate}
	\item Since the relevance score needs to be computed for millions of query-ad pairs in milliseconds, interaction based models are not feasible. In CLSM, the query and document (ad) embeddings are evaluated independently till the last layer. This implies that we can pre-calculate the embeddings for ads and need to compute only the query embedding at runtime to calculate the cosine similarity.

	\item The tri-letter representation makes it robust to out of vocabulary tokens and enables computation for tail queries as well.
\end{enumerate}
 
\subsection{CLSM}
CLSM aims to learn a low dimensional semantic vector representation of a query and a document to maximize the likelihood a positive document given the query. The semantic relevance of a query-document pair is calculated by taking the cosine similarity of their vector representations. This is converted to a posterior probability of the positive document given the query through softmax. CLSM has letter
trigram based word-n-gram representation. It performs a convolution over the word-n-gram representation followed by
max-pooling. The maxpooling layer is used to model sentence level semantic features. The output maxpool layer is
fed into a non-linear layer, whose matrix represents higher level latent semantic features. The CLSM represents the
relevance between query $q$ and document $d$ as
\[
R(q,d) = cos(y_q,y_d) = \frac{y_q^Ty_d}{||y_q||||y_d||}
\]
To train the model, CLSM suggested use of clicked data as positive data and randomly chosen pairs as negative data.

\subsection{Using CLSM to compute categorical similarity}
\label{CLSMCateg}
We modify the CLSM training data by defining positive data appropriately. Let
$g$ denote a function over query and ad pair. In our case, $g$ is a function that consumes query and ad categories,
and returns either a real number denoting category similarity. We denote set $D^{+}$ as a set of positive examples
such that
\[
D^{+} = \{(q, a) | g(q, a) > \delta\}
\]
where $\delta$ denotes a threshold above which we want to consider the pair as positive.

For the particular case of query-ad category match, we consider all the impressed query-ad pairs and consider those
pairs whose category match as the positive set. The objective of the learned CLSM with the modified positive data
generation scheme is to model whether the category of a query and an ad match or not. We call this model Category Match Approximator (CMA).

\subsection{Evaluation of CMA}
The training data for CMA is derived from the ad serving logs of a commercial search engine, over a period of 6 months. The serving logs provide details about query-ad pairs that were impressed and whether they received any clicks. The query and ad are annotated with categories from a 4000-category hierarchy maintained by the search engine. The query and ad are associated with top k categories based on the confidence score of a categorizer as in \cite{AdvIntent}. Since we are interested in modeling the category similarity between query and ad pair, we pick impressed query-ad pairs whose top category match to form the positive example. We use the query and ad title for training CMA model. We enforce that the top category should match exactly and don’t consider parent to descendant category match for positive examples. This is done with the motivation to capture both similar category and the same scope.
The query and ads are further preprocessed to be lower cased and additional whitespaces were removed. We also remove all non-alpha numeric characters. The training set consists of 19 million query-ad pairs and the evaluation consisted of 100,000 pairs generated using the process described above.

The first evaluation of the efficacy of CMA is its ability to correctly predict whether a given query-ad pair has the same top category or not. Since this is a binary classification problem, we use the standard metric of AUC-ROC to evaluate the effectiveness of the CMA. CMA has a ROC-AUC of 0.6758 on the evaluation set.
CMA tends to confuse between categories that have a close common ancestor. For a given query-ad pair, if the query’s (ad’s) category shares a close common ancestor with the ad’s (query’s) category, CMA predicts a higher score for them and hence incorrectly concludes that they belong to the same category. For example, The query "card printing adelaide" 
is assigned category \textit{Business Cards, Stationery \& Forms} that follows the path 
\textit{Business \& Industrial->Office->Business Cards, Stationery \& Forms}. The query, matched to an ad assigned category 
\textit{Commercial \& Industrial Printing} that follows the path \textit{Business \& Industrial->Commercial \& Industrial Printing} is given a high score.

The false negatives are caused in very wide top-level categories such as \textit{Books}. CMA generated false negatives when we had two books, but which were talking about completely different subjects. The false-positives
and false-negatives, although resulting in poor AUC-ROC, ultimately help in the final task. This is
because for cases where the top category is a generic category, predicting that the query and the ad belong to the same
category is semantically incorrect. Similarly, for some categories such as \textit{Italian Restaurant}
and \textit{Pizza Restaurant}, although there is a parent-child relationship, categorical equivalence can be considered in the overall task of relevance. Some examples where query-ad title pairs were labeled negative since their categories didn’t match exactly, but CMA gave high scores since they were categorically similar are given in table \ref{table:examples}.

\begin{table}
	\caption{Negatively labeled pairs having high CMA score}
	\begin{tabular}{ll}
		\textbf{Query} & \textbf{Ad title} \\
		\hline
		t shirt printer sale   & custom t shirt printing \\
		recover file in excel  & excel file recovery tool \\
		lower back exercise & back exercise lower back pain \\
	\end{tabular}
	\label{table:examples}
\end{table}
 For some cases, CMA predicted a high score for unrelated categories. The query-ad title pair “oral b brush”, “brush suppliers cost effective”, got a high score even though the ad is for industrial brush suppliers and is not categorically relevant to the query.

\subsection{Effect of Noise on CMA}
As described in section \ref{CLSMCateg}, the positive data for CMA are query-ad pairs whose top category match. In this section, we present the sensitivity of CMA results to the accuracy of the underlying categorizer. We modify the positive data generation process for CMA training. In addition to the generated positive examples, we also add impressed query-ad pairs which whose top category do not match. The motivation is to study the robustness of CMA learning with increasing proportion of incorrect positive examples. For different fraction of noisy pairs added, we retrained a new CMA and evaluated the model. The AUC-ROC corresponding to different models is given in table \ref{noise-auc}. CMA is robust to moderate fraction of noise in the positive samples. Beyond 0.3 fraction of noise, it loses its discriminative power.

\begin{table}
	\caption{AUC-ROC for different Fraction of Noise}
	\label{noise-auc}
	\begin{tabular}{ll}
		\textbf{Fraction of Actual Positive Points} & \textbf{AUC-ROC} \\
		\hline
		0.1 & $0.8164$ \\
		0.2 & $0.8092$ \\
		0.3 & $0.8011$ \\
		0.4 & $0.7962$ \\
	\end{tabular}
\end{table}

\section{Improving Sponsored Search Relevance}
\label{sec:expAdRel}
Showing relevant ads for queries is crucial to retain users, advertisers and for generating revenue through clicks. In a search engine, relevance evaluation occurs as a filtration step once ad matches for a query have been generated. 
Our query-ad relevance model is a Gradient Boosted Decision Tree (GBDT). Our model features are based on the 185 text-based features described in [2]. The features are generated between query and components of ad such as the bidded keyword, ad title, ad description, ad display URL and anchor texts from the ad’s landing page.

We considered three choices for the model:
\begin{enumerate}
 \item\textbf{Relevance-NoCat}: Category features are not used.
\item\textbf{Relevance-Binary}: Category features are represented as binary vector along with other features.
\item\textbf{Relevance-Derived}: Derived features such IsCategoryOverlapping, JaccardDistance based on the
  category vectors.
\item\textbf{Relevance-CMA}: CMA output is used as a feature
\end{enumerate}

To evaluate effectiveness of the CMA as a feature, we evaluated it against different choices of feature
representations. The evaluation set comprised of \textbf{77,000} query ad pairs sampled from the engine's
logs and hand labeled on a 5-level relevance scale: bad, fair, good, excellent and perfect where bad
corresponds to a completely irrelevant match and perfect corresponds to a highly relevant match. To obtain
the AUC-ROC, we merged bad and fair to label 0 and good, excellent, perfect to label 1. The comparison
of different feature representations are given in table \ref{table:relevance-auc}.

\begin{table}
  \caption{AUC-ROC for different Feature Representations}
  \begin{tabular}{ll}
    \textbf{Feature representation} & \textbf{AUC-ROC} \\
    \hline
    Relevance-NoCat   & $0.7972$ \\
    Relevance-Binary  & $0.7991$ \\
    Relevance-Derived & $0.8124$ \\
    Relevance-CMA    & $0.8389$ \\
  \end{tabular}
  \label{table:relevance-auc}
\end{table}

\begin{table}
	\caption{Query-ad pairs that were not blocked by Relevance-NoCat but were caught by Relevance-CMA}
	\begin{tabular}{ll}
		\textbf{Query} & \textbf{Ad title} \\
		\hline
		app won't download android & samsung android driver \\
		wine making equipment commercial
		  & commercial equipment leasing
		   \\
		product packaging waste australia & cd packaging australia \\
	\end{tabular}
	\label{table:nocat}
\end{table}

When we don’t use any category information in the relevance model, completely unrelated query-ad pairs are not always caught correctly. This is because lexical features alone have limited prediction capabilities when it comes to differentiating unrelated pairs and pairs that are semantically related but having little lexical overlap. Category information provides an informative prior, which is completely lacking when we don’t consider category features.
Relevance-Binary gives us a marginal improvement in AUC. Using binary feature vectors of category bloats the feature space and renders the model incapable of exploring the feature space fully with the provided number of training points. We see an improvement in the AUC by 1.9\% when derived categorical features are used. The improvement increases to 5.23\% when we use CMA to account for query-ad categorical similarity. Table   \ref{table:nocat} shows examples of query-ad pairs that were not caught by Relevance-NoCat but were blocked by Relevance-CMA.

\subsection{Online Experiments with CMA}
The relevance model with CMA was deployed on an actual commercial search engine to serve ads as part of A/B testing.The flight tests were conducted for a period of three weeks and tests for statistical significance were performed on the observed metrics. The model Relevance-Derived was used as the control model and Relevance-CMA as the treatment. To measure the improvement in the relevance of ads, a set of 10,000 queries was sampled to be indicative of the search engine’s query distribution. The ads impressed for this query set in the control and treatment were hand labeled based on whether the match was relevant or not. The treatment improved the relevance of query-ad pairs by 8.2\%. The treatment also saw an increase in CTR (click through rate) by 0.81\%.

\section{Conclusion}
We propose approximating categorical similarity between a query and an ad when modeling relevance in sponsored search. We use CLSM to approximate the similarity (CMA). The approach is robust to out-of-vocabulary tokens and allows computation of categories of tail queries at runtime. We show that incorporation of CMA as a feature in the relevance model instead of using categories directly shows significant gains in offline as well as online experiments.
\bibliographystyle{ACM-Reference-Format}
\bibliography{Paper_Bibliography}


\begin{thebibliography}{18}


\ifx \showCODEN    \undefined \def \showCODEN     #1{\unskip}     \fi
\ifx \showDOI      \undefined \def \showDOI       #1{#1}\fi
\ifx \showISBNx    \undefined \def \showISBNx     #1{\unskip}     \fi
\ifx \showISBNxiii \undefined \def \showISBNxiii  #1{\unskip}     \fi
\ifx \showISSN     \undefined \def \showISSN      #1{\unskip}     \fi
\ifx \showLCCN     \undefined \def \showLCCN      #1{\unskip}     \fi
\ifx \shownote     \undefined \def \shownote      #1{#1}          \fi
\ifx \showarticletitle \undefined \def \showarticletitle #1{#1}   \fi
\ifx \showURL      \undefined \def \showURL       {\relax}        \fi
\providecommand\bibfield[2]{#2}
\providecommand\bibinfo[2]{#2}
\providecommand\natexlab[1]{#1}
\providecommand\showeprint[2][]{arXiv:#2}

\bibitem[\protect\citeauthoryear{Aiello, Arapakis, Baeza-Yates, Bai, Barbieri,
  Mantrach, and Silvestri}{Aiello et~al\mbox{.}}{2016}]%
        {RoleRev}
\bibfield{author}{\bibinfo{person}{Luca Aiello}, \bibinfo{person}{Ioannis
  Arapakis}, \bibinfo{person}{Ricardo Baeza-Yates}, \bibinfo{person}{Xiao Bai},
  \bibinfo{person}{Nicola Barbieri}, \bibinfo{person}{Amin Mantrach}, {and}
  \bibinfo{person}{Fabrizio Silvestri}.} \bibinfo{year}{2016}\natexlab{}.
\newblock \showarticletitle{The Role of Relevance in Sponsored Search}. In
  \bibinfo{booktitle}{\emph{Proceedings of the 25th ACM International on
  Conference on Information and Knowledge Management}}
  \emph{(\bibinfo{series}{CIKM '16})}. \bibinfo{publisher}{ACM},
  \bibinfo{address}{New York, NY, USA}, \bibinfo{pages}{185--194}.
\newblock
\showISBNx{978-1-4503-4073-1}
\urldef\tempurl%
\url{https://doi.org/10.1145/2983323.2983840}
\showDOI{\tempurl}


\bibitem[\protect\citeauthoryear{Bai, Ordentlich, Zhang, Feng, Ratnaparkhi,
  Somvanshi, and Tjahjadi}{Bai et~al\mbox{.}}{2018}]%
        {NGram}
\bibfield{author}{\bibinfo{person}{Xiao Bai}, \bibinfo{person}{Erik
  Ordentlich}, \bibinfo{person}{Yuanyuan Zhang}, \bibinfo{person}{Andy Feng},
  \bibinfo{person}{Adwait Ratnaparkhi}, \bibinfo{person}{Reena Somvanshi},
  {and} \bibinfo{person}{Aldi Tjahjadi}.} \bibinfo{year}{2018}\natexlab{}.
\newblock \showarticletitle{Scalable Query N-Gram Embedding for Improving
  Matching and Relevance in Sponsored Search}. In
  \bibinfo{booktitle}{\emph{Proceedings of the 24th ACM SIGKDD International
  Conference on Knowledge Discovery \& Data Mining}}. ACM,
  \bibinfo{pages}{52--61}.
\newblock


\bibitem[\protect\citeauthoryear{Bauer, Radlinski, and White}{Bauer
  et~al\mbox{.}}{2016}]%
        {Boulder}
\bibfield{author}{\bibinfo{person}{Sandro Bauer}, \bibinfo{person}{Filip
  Radlinski}, {and} \bibinfo{person}{Ryen~W White}.}
  \bibinfo{year}{2016}\natexlab{}.
\newblock \showarticletitle{Where Can I Buy a Boulder?: Searching for Offline
  Retail Locations}. In \bibinfo{booktitle}{\emph{Proceedings of the 25th
  International Conference on World Wide Web}}. International World Wide Web
  Conferences Steering Committee, \bibinfo{pages}{1225--1235}.
\newblock


\bibitem[\protect\citeauthoryear{Blei, Ng, and Jordan}{Blei
  et~al\mbox{.}}{2003}]%
        {LDA}
\bibfield{author}{\bibinfo{person}{David~M Blei}, \bibinfo{person}{Andrew~Y
  Ng}, {and} \bibinfo{person}{Michael~I Jordan}.}
  \bibinfo{year}{2003}\natexlab{}.
\newblock \showarticletitle{Latent dirichlet allocation}.
\newblock \bibinfo{journal}{\emph{Journal of machine Learning research}}
  \bibinfo{volume}{3}, \bibinfo{number}{Jan} (\bibinfo{year}{2003}),
  \bibinfo{pages}{993--1022}.
\newblock


\bibitem[\protect\citeauthoryear{Broder, Fontoura, Gabrilovich, Joshi,
  Josifovski, and Zhang}{Broder et~al\mbox{.}}{2007}]%
        {Class}
\bibfield{author}{\bibinfo{person}{Andrei~Z Broder}, \bibinfo{person}{Marcus
  Fontoura}, \bibinfo{person}{Evgeniy Gabrilovich}, \bibinfo{person}{Amruta
  Joshi}, \bibinfo{person}{Vanja Josifovski}, {and} \bibinfo{person}{Tong
  Zhang}.} \bibinfo{year}{2007}\natexlab{}.
\newblock \showarticletitle{Robust classification of rare queries using web
  knowledge}. In \bibinfo{booktitle}{\emph{Proceedings of the 30th annual
  international ACM SIGIR conference on Research and development in information
  retrieval}}. ACM, \bibinfo{pages}{231--238}.
\newblock


\bibitem[\protect\citeauthoryear{Deerwester, Dumais, Furnas, Landauer, and
  Harshman}{Deerwester et~al\mbox{.}}{1990}]%
        {LSA}
\bibfield{author}{\bibinfo{person}{Scott Deerwester}, \bibinfo{person}{Susan~T
  Dumais}, \bibinfo{person}{George~W Furnas}, \bibinfo{person}{Thomas~K
  Landauer}, {and} \bibinfo{person}{Richard Harshman}.}
  \bibinfo{year}{1990}\natexlab{}.
\newblock \showarticletitle{Indexing by latent semantic analysis}.
\newblock \bibinfo{journal}{\emph{Journal of the American society for
  information science}} \bibinfo{volume}{41}, \bibinfo{number}{6}
  (\bibinfo{year}{1990}), \bibinfo{pages}{391--407}.
\newblock


\bibitem[\protect\citeauthoryear{Downey, Dumais, and Horvitz}{Downey
  et~al\mbox{.}}{2007}]%
        {Tail}
\bibfield{author}{\bibinfo{person}{Doug Downey}, \bibinfo{person}{Susan
  Dumais}, {and} \bibinfo{person}{Eric Horvitz}.}
  \bibinfo{year}{2007}\natexlab{}.
\newblock \showarticletitle{Heads and Tails: Studies of Web Search with Common
  and Rare Queries}. In \bibinfo{booktitle}{\emph{Proceedings of the 30th
  Annual International ACM SIGIR Conference on Research and Development in
  Information Retrieval}} \emph{(\bibinfo{series}{SIGIR '07})}.
  \bibinfo{publisher}{ACM}, \bibinfo{address}{New York, NY, USA},
  \bibinfo{pages}{847--848}.
\newblock
\showISBNx{978-1-59593-597-7}
\urldef\tempurl%
\url{https://doi.org/10.1145/1277741.1277939}
\showDOI{\tempurl}


\bibitem[\protect\citeauthoryear{Grbovic, Djuric, Radosavljevic, Silvestri,
  Baeza-Yates, Feng, Ordentlich, Yang, and Owens}{Grbovic
  et~al\mbox{.}}{2016}]%
        {SemMatch}
\bibfield{author}{\bibinfo{person}{Mihajlo Grbovic}, \bibinfo{person}{Nemanja
  Djuric}, \bibinfo{person}{Vladan Radosavljevic}, \bibinfo{person}{Fabrizio
  Silvestri}, \bibinfo{person}{Ricardo Baeza-Yates}, \bibinfo{person}{Andrew
  Feng}, \bibinfo{person}{Erik Ordentlich}, \bibinfo{person}{Lee Yang}, {and}
  \bibinfo{person}{Gavin Owens}.} \bibinfo{year}{2016}\natexlab{}.
\newblock \showarticletitle{Scalable Semantic Matching of Queries to Ads in
  Sponsored Search Advertising}. In \bibinfo{booktitle}{\emph{Proceedings of
  the 39th International ACM SIGIR Conference on Research and Development in
  Information Retrieval}} \emph{(\bibinfo{series}{SIGIR '16})}.
  \bibinfo{publisher}{ACM}, \bibinfo{address}{New York, NY, USA},
  \bibinfo{pages}{375--384}.
\newblock
\showISBNx{978-1-4503-4069-4}
\urldef\tempurl%
\url{https://doi.org/10.1145/2911451.2911538}
\showDOI{\tempurl}


\bibitem[\protect\citeauthoryear{Hillard, Schroedl, Manavoglu, Raghavan, and
  Leggetter}{Hillard et~al\mbox{.}}{2010}]%
        {ImproveRel}
\bibfield{author}{\bibinfo{person}{Dustin Hillard}, \bibinfo{person}{Stefan
  Schroedl}, \bibinfo{person}{Eren Manavoglu}, \bibinfo{person}{Hema Raghavan},
  {and} \bibinfo{person}{Chirs Leggetter}.} \bibinfo{year}{2010}\natexlab{}.
\newblock \showarticletitle{Improving Ad Relevance in Sponsored Search}. In
  \bibinfo{booktitle}{\emph{Proceedings of the Third ACM International
  Conference on Web Search and Data Mining}} \emph{(\bibinfo{series}{WSDM
  '10})}. \bibinfo{publisher}{ACM}, \bibinfo{address}{New York, NY, USA},
  \bibinfo{pages}{361--370}.
\newblock
\showISBNx{978-1-60558-889-6}
\urldef\tempurl%
\url{https://doi.org/10.1145/1718487.1718532}
\showDOI{\tempurl}


\bibitem[\protect\citeauthoryear{Hofmann}{Hofmann}{2017}]%
        {pLSI}
\bibfield{author}{\bibinfo{person}{Thomas Hofmann}.}
  \bibinfo{year}{2017}\natexlab{}.
\newblock \showarticletitle{Probabilistic latent semantic indexing}. In
  \bibinfo{booktitle}{\emph{ACM SIGIR Forum}}, Vol.~\bibinfo{volume}{51}. ACM,
  \bibinfo{pages}{211--218}.
\newblock


\bibitem[\protect\citeauthoryear{Jiang, Hu, Kang, Daly, Yin, Chang, and
  Zhai}{Jiang et~al\mbox{.}}{2016}]%
        {WebClick}
\bibfield{author}{\bibinfo{person}{Shan Jiang}, \bibinfo{person}{Yuening Hu},
  \bibinfo{person}{Changsung Kang}, \bibinfo{person}{Tim Daly, Jr.},
  \bibinfo{person}{Dawei Yin}, \bibinfo{person}{Yi Chang}, {and}
  \bibinfo{person}{Chengxiang Zhai}.} \bibinfo{year}{2016}\natexlab{}.
\newblock \showarticletitle{Learning Query and Document Relevance from a
  Web-scale Click Graph}. In \bibinfo{booktitle}{\emph{Proceedings of the 39th
  International ACM SIGIR Conference on Research and Development in Information
  Retrieval}} \emph{(\bibinfo{series}{SIGIR '16})}. \bibinfo{publisher}{ACM},
  \bibinfo{address}{New York, NY, USA}, \bibinfo{pages}{185--194}.
\newblock
\showISBNx{978-1-4503-4069-4}
\urldef\tempurl%
\url{https://doi.org/10.1145/2911451.2911531}
\showDOI{\tempurl}


\bibitem[\protect\citeauthoryear{Liu, Chang, Wu, and Yang}{Liu
  et~al\mbox{.}}{2017}]%
        {FastText}
\bibfield{author}{\bibinfo{person}{Jingzhou Liu}, \bibinfo{person}{Wei-Cheng
  Chang}, \bibinfo{person}{Yuexin Wu}, {and} \bibinfo{person}{Yiming Yang}.}
  \bibinfo{year}{2017}\natexlab{}.
\newblock \showarticletitle{Deep learning for extreme multi-label text
  classification}. In \bibinfo{booktitle}{\emph{Proceedings of the 40th
  International ACM SIGIR Conference on Research and Development in Information
  Retrieval}}. ACM, \bibinfo{pages}{115--124}.
\newblock


\bibitem[\protect\citeauthoryear{Mikolov, Chen, Corrado, and Dean}{Mikolov
  et~al\mbox{.}}{2013a}]%
        {SNGS}
\bibfield{author}{\bibinfo{person}{Tomas Mikolov}, \bibinfo{person}{Kai Chen},
  \bibinfo{person}{Greg Corrado}, {and} \bibinfo{person}{Jeffrey Dean}.}
  \bibinfo{year}{2013}\natexlab{a}.
\newblock \showarticletitle{Efficient Estimation of Word Representations in
  Vector Space}.
\newblock \bibinfo{journal}{\emph{CoRR}}  \bibinfo{volume}{abs/1301.3781}
  (\bibinfo{year}{2013}).
\newblock
\showeprint[arxiv]{1301.3781}
\urldef\tempurl%
\url{http://arxiv.org/abs/1301.3781}
\showURL{%
\tempurl}


\bibitem[\protect\citeauthoryear{Mikolov, Sutskever, Chen, Corrado, and
  Dean}{Mikolov et~al\mbox{.}}{2013b}]%
        {CBOW}
\bibfield{author}{\bibinfo{person}{Tomas Mikolov}, \bibinfo{person}{Ilya
  Sutskever}, \bibinfo{person}{Kai Chen}, \bibinfo{person}{Greg Corrado}, {and}
  \bibinfo{person}{Jeffrey Dean}.} \bibinfo{year}{2013}\natexlab{b}.
\newblock \showarticletitle{Distributed Representations of Words and Phrases
  and their Compositionality}.
\newblock \bibinfo{journal}{\emph{CoRR}}  \bibinfo{volume}{abs/1310.4546}
  (\bibinfo{year}{2013}).
\newblock
\showeprint[arxiv]{1310.4546}
\urldef\tempurl%
\url{http://arxiv.org/abs/1310.4546}
\showURL{%
\tempurl}


\bibitem[\protect\citeauthoryear{Mitra, Diaz, and Craswell}{Mitra
  et~al\mbox{.}}{2017}]%
        {ContextEmbed}
\bibfield{author}{\bibinfo{person}{Bhaskar Mitra}, \bibinfo{person}{Fernando
  Diaz}, {and} \bibinfo{person}{Nick Craswell}.}
  \bibinfo{year}{2017}\natexlab{}.
\newblock \showarticletitle{Learning to match using local and distributed
  representations of text for web search}. In
  \bibinfo{booktitle}{\emph{Proceedings of the 26th International Conference on
  World Wide Web}}. International World Wide Web Conferences Steering
  Committee, \bibinfo{pages}{1291--1299}.
\newblock


\bibitem[\protect\citeauthoryear{Sculley, Malkin, Basu, and Bayardo}{Sculley
  et~al\mbox{.}}{2009}]%
        {Bounce}
\bibfield{author}{\bibinfo{person}{D Sculley}, \bibinfo{person}{Robert~G
  Malkin}, \bibinfo{person}{Sugato Basu}, {and} \bibinfo{person}{Roberto~J
  Bayardo}.} \bibinfo{year}{2009}\natexlab{}.
\newblock \showarticletitle{Predicting bounce rates in sponsored search
  advertisements}. In \bibinfo{booktitle}{\emph{Proceedings of the 15th ACM
  SIGKDD international conference on Knowledge discovery and data mining}}.
  ACM, \bibinfo{pages}{1325--1334}.
\newblock


\bibitem[\protect\citeauthoryear{Shen, He, Gao, Deng, and Mesnil}{Shen
  et~al\mbox{.}}{2014}]%
        {CLSM}
\bibfield{author}{\bibinfo{person}{Yelong Shen}, \bibinfo{person}{Xiaodong He},
  \bibinfo{person}{Jianfeng Gao}, \bibinfo{person}{Li Deng}, {and}
  \bibinfo{person}{Gr{\'e}goire Mesnil}.} \bibinfo{year}{2014}\natexlab{}.
\newblock \showarticletitle{A Latent Semantic Model with Convolutional-Pooling
  Structure for Information Retrieval}. In
  \bibinfo{booktitle}{\emph{Proceedings of the 23rd ACM International
  Conference on Conference on Information and Knowledge Management}}
  \emph{(\bibinfo{series}{CIKM '14})}. \bibinfo{publisher}{ACM},
  \bibinfo{address}{New York, NY, USA}, \bibinfo{pages}{101--110}.
\newblock
\showISBNx{978-1-4503-2598-1}
\urldef\tempurl%
\url{https://doi.org/10.1145/2661829.2661935}
\showDOI{\tempurl}


\bibitem[\protect\citeauthoryear{Vattikonda, Kodipaka, Zhou, Dave, Guha, and
  Snoeren}{Vattikonda et~al\mbox{.}}{2015}]%
        {AdvIntent}
\bibfield{author}{\bibinfo{person}{Bhanu~C Vattikonda},
  \bibinfo{person}{Santhosh Kodipaka}, \bibinfo{person}{Hongyan Zhou},
  \bibinfo{person}{Vacha Dave}, \bibinfo{person}{Saikat Guha}, {and}
  \bibinfo{person}{Alex~C Snoeren}.} \bibinfo{year}{2015}\natexlab{}.
\newblock \showarticletitle{Interpreting advertiser intent in sponsored
  search}. In \bibinfo{booktitle}{\emph{Proceedings of the 21th ACM SIGKDD
  International Conference on Knowledge Discovery and Data Mining}}. ACM,
  \bibinfo{pages}{2177--2185}.
\newblock


\end{thebibliography}

\end{document}